\documentclass[aps,prl,twocolumn,superscriptaddress,a4paper]{revtex4}

\usepackage{graphicx,dcolumn,bm,url}
\usepackage[colorlinks=true,linkcolor=red,citecolor=blue]{hyperref}

\newcommand{\be}{\begin{equation}}
\newcommand{\ee}{\end{equation}}
\newcommand{\ben}{\begin{eqnarray}}
\newcommand{\een}{\end{eqnarray}}
\newcommand{\bes}{\begin{subequations}}
\newcommand{\ees}{\end{subequations}}
\newcommand{\bF}{\begin{figure}}
\newcommand{\eF}{\end{figure}}

\newcommand{\arxiv}[2][arxiv:]{\href{http://arxiv.org/abs/#1#2}{#1#2}}

\def\tr{ {\rm{Tr }}\,}
\newcommand{\proj}[1]{\mbox{$|#1\rangle \!\langle #1 |$}}

\newcommand{\ket}[1]{\ensuremath| #1 \rangle}
\newcommand{\bra}[1]{\ensuremath\langle #1 |}

\def\tr{ {\rm{Tr }}}

\begin{document}

\title{Mapping coherence in measurement via full quantum tomography of a hybrid optical detector}

\author{Lijian Zhang$^*$}
\affiliation{Clarendon Laboratory, Department of Physics, University of Oxford, OX1 3PU, United Kingdom}

\author{Hendrik Coldenstrodt-Ronge}
\affiliation{Clarendon Laboratory, Department of Physics, University of Oxford, OX1 3PU, United Kingdom}

\author{Animesh Datta}
\affiliation{Clarendon Laboratory, Department of Physics, University of Oxford, OX1 3PU, United Kingdom}

\author{Graciana Puentes}
\affiliation{Department of Physics, Massachusetts Institute of Technology, Cambridge MA 02139, USA}

\author{Jeff S. Lundeen}
\affiliation{Institute for National Measurement Standards, National Research Council, Montreal Road, Ottawa, K1A 0R6, Canada}

\author{Xian-Min Jin}
\affiliation{Clarendon Laboratory, Department of Physics, University of Oxford, OX1 3PU, United Kingdom}
\affiliation{Centre for Quantum Technologies, National University of Singapore, Singapore}

\author{Brian J. Smith}
\affiliation{Clarendon Laboratory, Department of Physics, University of Oxford, OX1 3PU, United Kingdom}

\author{Martin B. Plenio}
\affiliation{Institut f\"{u}r Theoretische Physik, Albert-Einstein-Allee 11, Universit\"{a}t Ulm, D-89069 Ulm, Germany}
\affiliation{QOLS, Blackett Laboratory, Imperial College London, SW7 2BW, United Kingdom}

\author{Ian A. Walmsley}
\affiliation{Clarendon Laboratory, Department of Physics, University of Oxford, OX1 3PU, United Kingdom}

\maketitle

{ \bf Quantum states and measurements exhibit wave-like --- continuous, or particle-like --- discrete, character. Hybrid discrete-continuous photonic systems are key to investigating fundamental quantum phenomena~\cite{Kuzmich00, SchukinVogel06, Parigi2007}, 
generating superpositions of macroscopic states~\cite{Ourjoumtsev07}, and form essential resources for quantum-enhanced applications~\cite{BraunsteinvanLoock05}, \textit{e.g.} entanglement distillation~\cite{ourjoumtsev-2007-98, takahashi2010entanglement} and quantum computation~\cite{Sasaki10}, as well as highly efficient optical telecommunications~\cite{Guha11,Kenji11}. Realizing the full potential of these hybrid systems requires quantum-optical measurements sensitive to complementary observables such as field quadrature amplitude and photon number~\cite{Puentes09,Lvovsky10,Laiho_CGS10}. However, a thorough understanding of the practical performance of an optical detector interpolating between these two regions is absent. Here, we report the implementation of full quantum detector tomography, enabling the characterization of the simultaneous wave and photon-number sensitivities of quantum-optical detectors. This yields the largest parametrization to-date in quantum tomography experiments, requiring the development of novel theoretical tools. Our results reveal the role of coherence in quantum measurements and demonstrate the tunability of hybrid quantum-optical detectors.}

Accurate knowledge of a quantum-optical detector is essential for its fruitful utilization, be it in foundational investigations or technological applications. Photodetectors are normally characterized by several parameters, including detectivity, spectral sensitivity and noise-equivalent power~\cite{boyd1983radiometry}. For quantum detectors, additional information is required for a complete specification of the detector. This information is the set of operators that link the input quantum state of the light field to the classical detector output, known as postive-operator-valued measure (POVM). It may be estimated by means of quantum detector tomogrpahy (QDT)~\cite{Luis99,ML01,Ariano04,lundeen2008tomography}, and is needed if the detector is to be used reliably. To date, QDT has been successfully applied to avalanche photodiodes (APDs)~\cite{Fabre11}, time-multiplexed detectors~\cite{lundeen2008tomography,feito2009measuring,coldenstrodt2009proposed}, transition-edge sensors~\cite{Brida11}, and super-conducting nanowire detectors~\cite{nanotomo}. The matrix representations of the POVMs for these detectors are diagonal in the photon-number basis. Consequently the reconstruction problem is linear and positive, and therefore amenable to solution by means familiar to classical signal processing~\cite{Hradil01}. This is not true for a general quantum detector: the POVM elements can have non-zero off-diagonals due to coherent superpositions. Even in conventional optical communications, coherent modulation and detection can increase the data transmission rate by an order of magnitude. Moreover, exploration and utilization of the full Hilbert space of a quantum system requires a detector capable of implementing a tomographically complete set of measurements~\cite{Nunn10}. Such a capability is also vital to fully harness the potential of hybrid quantum systems operating at the confluence of discrete and continuous variable regimes. To this end, phase-sensitive detectors that can measure coherent superpositions of photon-number states are essential to both quantum and classical optical applications. We focus on the particularly interesting example of the weak-homodyne photon detector, which allows us to navigate the discrete-continuous confluence region in a tunable manner.

\begin{figure}[ht]
\includegraphics[width=.45\textwidth]{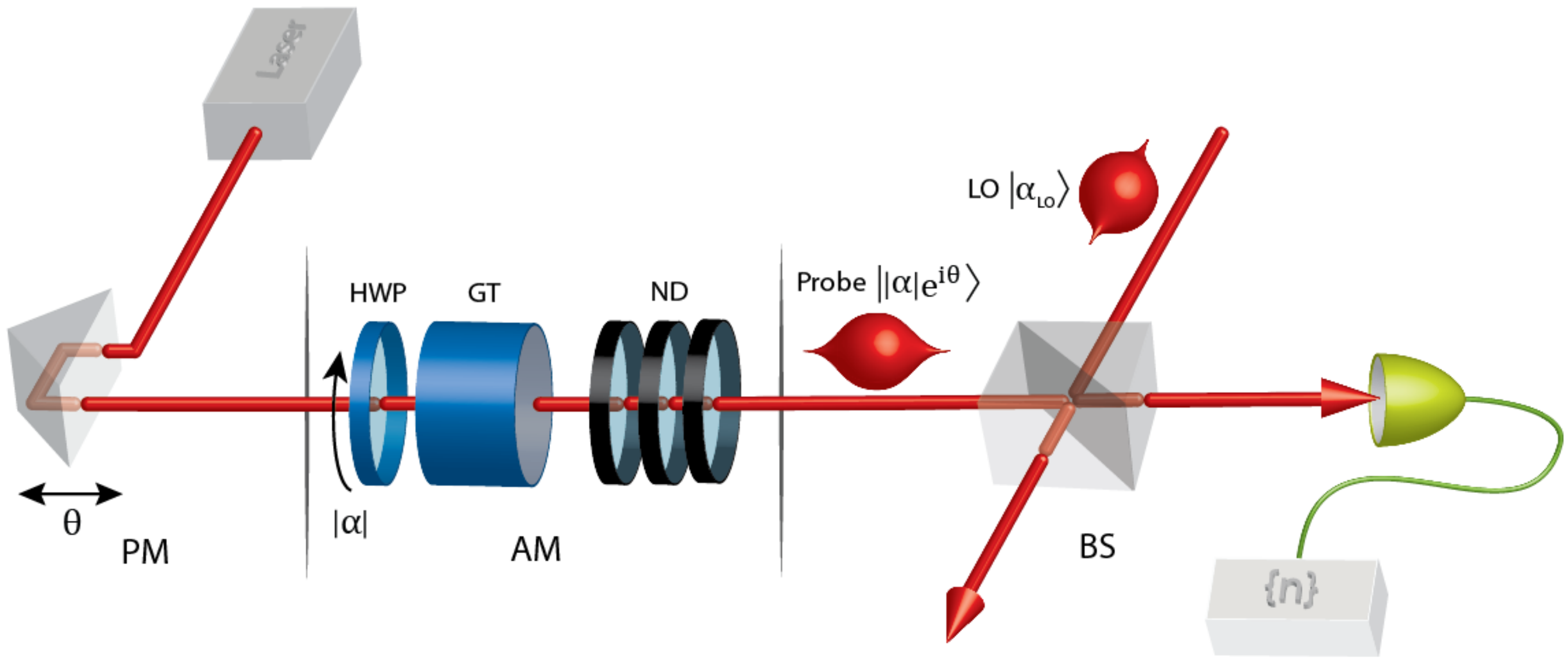}
\caption{Experimental setup. The output of a laser goes through a phase modulator (PM) and an amplitude modulator (AM) to prepare a set of probe states which are injected into the detector. The magnitudes of the probe states $|\alpha|$ are controlled by a motorized half-wave plate followed by a Glan-Thompson polarizer. The phases of the probe states $\theta$ are set by a piezo translator (see Methods: Experimental details). From the measurement statistics, the detector POVM elements are reconstructed. HWP: half-wave plate. GT: Glan-Thompson polarizer. ND: neutral density filter. LO: local oscillator. BS: beam splitter.}
\label{Fig:setup}
\end{figure}

In this Letter, we introduce a QDT method for the reconstruction of the POVM of a coherent optical detector. This method is applied to two variants of a weak-homodyne detector: photon-counting and photon-number-resolving (PNRD). The POVM elements of such detectors have both phase and number sensitivity, denoted by their off-diagonal and diagonal matrix elements respectively. Our experimental procedure, shown schematically in Fig.~(\ref{Fig:setup}), can be universally applied to any optical detector. It uses only classical optical states as probes, yet, with the resulting POVM we can predict the detector response to any quantum state, including non-classical ones. Full quantum tomography is only realized by the development of a new recursive algorithm that radically reduces the computational complexity of reconstructing the POVM. The new recipe changes the complexity from quadratic to linear per recursion in the dimension $d$ of the POVM elements. This enables us to reconstruct a matrix of unprecedented size representing a quantum operation, essential for situations where the size of the Hilbert space is incompletely known. In particular, we reconstruct a POVM with $1.8\times 10^6$ parameters, almost two orders of larger than the largest quantum tomography ever performed. By defining the transformed version of the Husimi distribution ($Q$ function), we cast the reconstruction problem as a tractable semi-definite program, allowing us to determine both diagonal and off-diagonal elements. This enables us to characterize the detectors' tunability between registering the particle and wave behaviors of an input quantum state, making our experiment the first full QDT. This also allows us to completely characterize phase-sensitive optical detectors.

QDT is performed by preparing a set of known probe states $\{\hat{\rho}_m\}$ incident on a quantum detector and observing the detector outcomes. The probability of registering outcome $n$ is given by the Born rule
\begin{equation}
p_{n|m} = \tr(\hat{\rho}_m \hat{\Pi}_{n}),
\label{eq:Born}
\end{equation}
where $\{\hat{\Pi}_{n}\}$ is the POVM of the detector. The POVM elements can be reconstructed from these equations using convex optimization. Here we use coherent states as probes since they form a tomographically complete set. The set of measured outcome statistics gives the Husimi $Q$ function of the detector operator $Q_n(\alpha) = \bra{\alpha}\hat{\Pi}_{n}\ket{\alpha}/\pi$, where $\alpha = |\alpha|e^{i\theta}$ is the complex amplitude of the coherent probe state. $Q_n(\alpha)$ contains all the information about the detector, and can be used to make any predictions of measurement outcomes. In particular, all detection probabilities are given by the overlap of this $Q$ function with the Glauber-Sudarshan $P$ function of the quantum state incident upon the detector~\cite{Hillery_OSW84}. As the $P$-function of a non-classical state is highly singular, typically involving derivatives of the Dirac-delta function, it makes the overlap extremely sensitive to noise in $Q_n(\alpha)$, rendering its use impossible in practice. It is thus necessary to reconstruct the POVM elements $\{\hat{\Pi}_{n}\}$ themselves.

For optical detectors, POVM elements can be written in the photon-number basis as $\hat{\Pi}_{n} = \sum_{j,k}\Pi_{n}^{j,k}\ket{j}\bra{k}$. Detector saturation allows truncation of the Hilbert space at a finite number of $d-1$ photons, leaving us with $d^2-1$ parameters to estimate for each POVM element. In quantum state tomography, one can reconstruct the full density matrix because $d$ is typically small due to the lack of bright quantum-optical sources. In contrast, quantum-optical detectors can have a large dynamic range, with $d$ typically in the range of order $10^2$ to $10^5$. This makes reconstructing $\mathcal{O}(d^2)$ parameters extremely challenging. We overcome this problem by using the transformed version of the Husimi
distribution
\begin{equation}
\int_{0}^{2\pi} Q_n(\alpha)e^{-il\theta} d\theta = 2e^{-|\alpha|^2}\sum_{j} \Pi_{n}^{j,j+l} \frac{|\alpha|^{2j+l}}{\sqrt{(j+l)!j!}}.
\label{eq:rela_q_diag}
\end{equation}
This reduced function enables the recursive reconstruction of the principle diagonal ($l=0$) and then each leading off-diagonal ($l=1\cdots d$) (See Methods: Reconstruction procedure). The number of coefficients calculated per POVM element per recursion is now no more than $d$, greatly reducing the complexity. For instance, $l=0$ describes a phase-averaged coherent state as input, for which the detection probabilities involve only the principle diagonals of the POVM elements. Although one could reconstruct the entire operator using Eq.~(\ref{eq:rela_q_diag}), in many situations, losses and phase fluctuations restrict the number of significant off-diagonals to $l\ll d$. Moreover, situations involving input states with a fixed photon number $N$, like $N00N$ states~\cite{Sanders_89} or Holland-Burnett states~\cite{Datta_ZT-PDSW11}, require only $N$ leading diagonals of the POVM elements to predict all measurement outcomes. 

We apply the above strategy to the tomography of a weak-homodyne detector. This detector combines the input state with a local oscillator (LO), typically a weak coherent state $\ket{\alpha_{LO}}$ at a beam splitter (BS) with transmissivity $T$ ($65.5\%$ in our experiment), followed by an APD or a PNRD. The LO provides a phase reference to access coherence within the input states through interference at the BS. This enables the detector to probe both the particle and wave nature of a quantum state by tuning $|\alpha_{LO}|$~\cite{Puentes09}. Theoretical models of these detectors are affected by the BS ratio, LO amplitude and phase, as well as the detailed workings of the APD or PNRD. Additional factors such as the degree of mode overlap $\mathcal{M}$ between the input states and the LO and different losses of the input states and the LO, which are difficult to measure accurately, must be incorporated into the theoretical model. This makes an empirical characterization of these detectors, such as QDT, more rigorous and reliable, thus revealing their tunable quantum features. 

\begin{figure*}[t!]
\centerline{\includegraphics[width=0.9\textwidth]{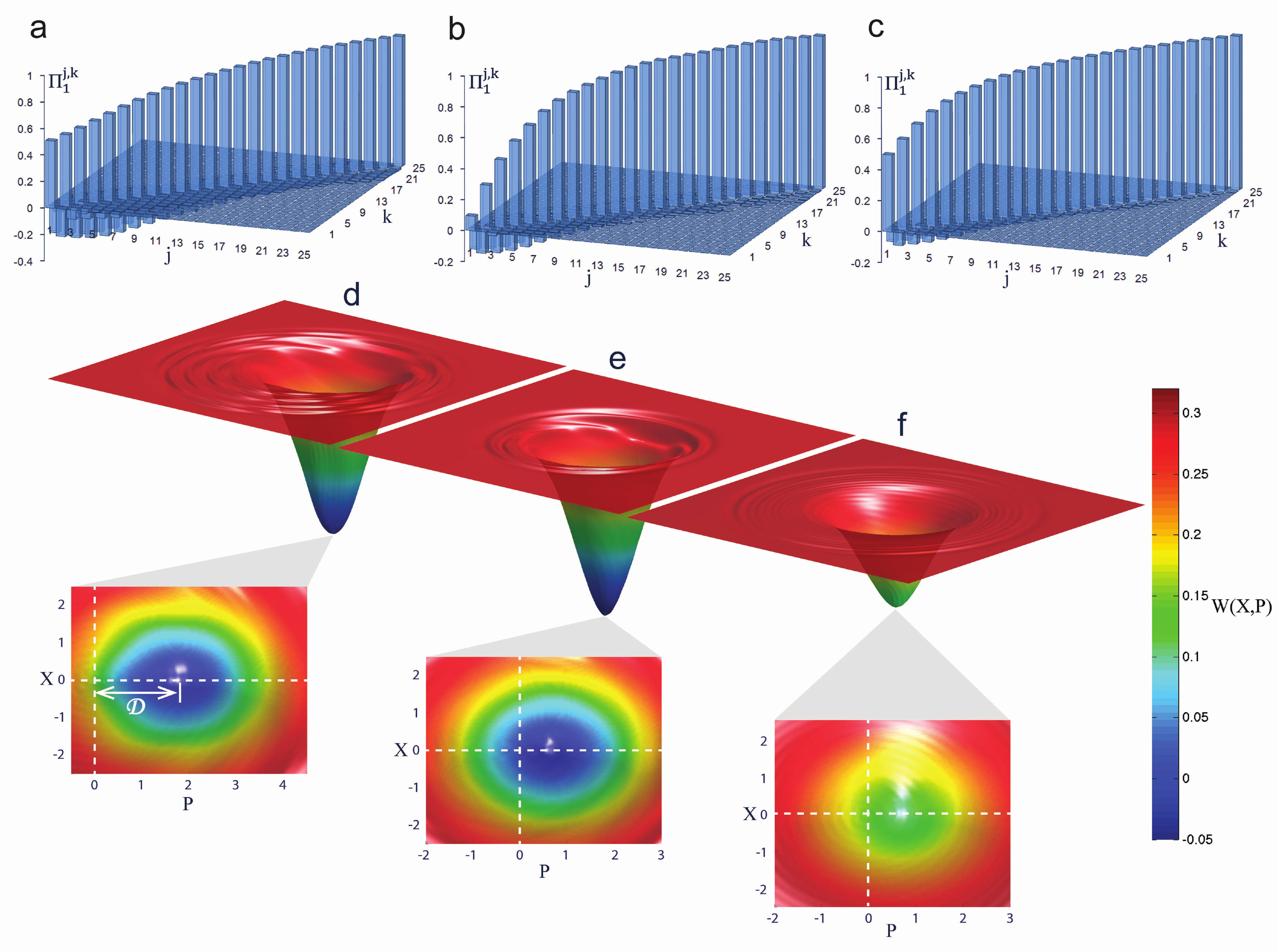}}
\caption{Experimentally reconstructed 1-click POVM elements of a weak-homodyne APD. Left to right: LO strengths $|\alpha_{LO}|^2 = 5.5, 0.8, 5.5$ photons, and a mode overlap $\mathcal{M} = 0.99, 0.99, 0.16$ respectively. Only the relative phase between the input states and LO determines the measurement result. We assume the LO phase to be 0, making the matrix elements all real. \textbf{Top}: (a)-(c): POVM elements in Fock basis. Decreased effective LO strength $\mathcal{M}|\alpha_{LO}|^2$ leads to suppressed off-diagonal elements of the POVM matrix, which represents the coherence in the detector, clearly demonstrating the ability of the weak-homodyne detector to bridge the wave-particle gap through a tunable LO. The off-diagonal elements in (c) are more akin to those in (b) than (a) due to the effective LO. The fidelities of the reconstructed POVMs $\Pi_n^{\mathrm{rec}}$ with the theoretical predictions $\Pi_n^{\mathrm{th}},$ given by $\mathcal{F}_n =\left(\tr\sqrt{\sqrt{\Pi_n^{\mathrm{th}}} \Pi_n^{\mathrm{rec}}\sqrt{\Pi_n^{\mathrm{th}}}}\right)^2/\tr(\Pi_n^{\mathrm{th}}) \tr(\Pi_n^{\mathrm{rec}}) $ are over $98\%$. \textbf{Bottom}: (d)-(f): Wigner functions of the above POVMs. The measured displacements $\mathcal{D}$ from the origin are 1.76, 0.56 and 0.64, and the values calculated from Eq.~(\ref{eq:displacement}) are 1.70, 0.65 and 0.68. The discrepency between experimental results and theoretical predictions can be explained by limited precisions in the measurement of the experimental parameters ($T$, $\mathcal{M}$ and $|\alpha_{LO}|$), as well as the noise in the reconstructed Wigner functions which limits the accuracy of estimating $\mathcal{D}$.}
\label{Fig:homo_APD}
\end{figure*}

We first explore experimentally the role of coherence in quantum measurements performed by a weak-homodyne APD (detection efficiency $39\%$, measured with a separate QDT) by modulating $|\alpha_{LO}|$ and $\mathcal{M}$. The resulting variations in the relative magnitudes of the diagonals and off-diagonals of the POVM elements govern the number and phase sensitivities of the detector, as in Fig.~(\ref{Fig:homo_APD}~a-c). When the detection efficiency is fixed, for perfect mode-overlap as in Fig.~(\ref{Fig:homo_APD}~a,b), reducing the LO strength leads to suppressed phase sensitivity. Mode-mismatch leads to two incoherent processes: (i) The interference of the reduced LO with the input state, the former with a strength $\mathcal{M}|\alpha_{LO}|^2$. (ii) Mixing of the remainder of the LO, $(1-\mathcal{M})|\alpha_{LO}|^2$, with vacuum at the input. Process (i) reduces the phase sensitivity, and process (ii) has no phase sensitivity. The total response of the detector is the convolution of the above two processes. From this perspective, we can explain why the off-diagonals in Figs.~(\ref{Fig:homo_APD}~b) and (c) are similar. The different LO strength between the two cases is offset by their different mode overlaps, resulting in similar effective LO strengths $\mathcal{M}|\alpha_{LO}|^2$. The convolution also leads to a modulation in the diagonals. This behavior is present in the different diagonals of Figs.~(\ref{Fig:homo_APD}~b) and (c).

\begin{figure*}[ht]
\includegraphics[width=0.9\textwidth]{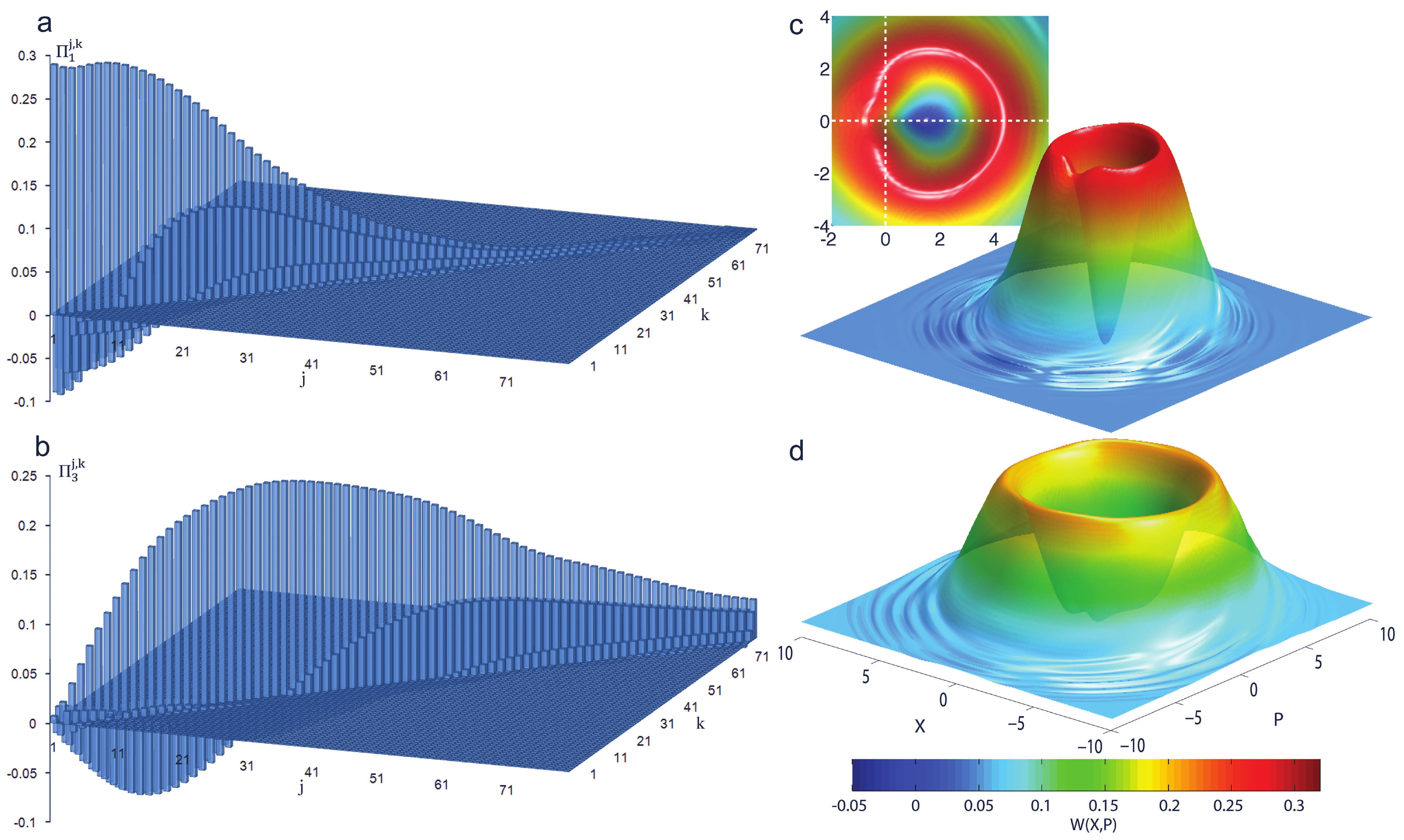}
\caption{Reconstructed POVM elements for (a) 1-click and (b) 3-click events of the weak-homodyne time-multiplexed PNRD in Fock basis. Again the LO phase is 0. We reconstruct up to the 6 leading off-diagonals, beyond which the coefficients are negligible. The corresponding Wigner functions are shown in (c) and (d). The simultaneous phase and number sensitivity is apparent from the radial asymmetry of the Wigner functions, as is the non-classicality of the operator, from their negativity. The 1-click Wigner function has an overlap of $98\%$ with a displaced single-photon state (having experienced $84.3\%$ loss).}
 \label{Fig:3click}
\end{figure*}

Our results can be further elucidated through the Wigner functions of the POVM elements in Fig.~(\ref{Fig:homo_APD} d - f). The presence of off-diagonals in an operator leads to radial asymmetry in its Wigner function. Since the response of a detector to an input state is determined by the overlap of their respective Wigner functions, the phase sensitivity of a POVM element can be inferred immediately from the radial asymmetry of its Wigner function. At one extreme is the Wigner function corresponding to a detection outcome for a standard homodyne detector (i.e. one in which the LO is strong and the detectors are linear photodiodes), which is a delta plane in the phase space. The opposite extreme is the Wigner function of a Fock state projection, a radial annulus. The experimentally reconstructed Wigner functions of our weak-homodyne detectors demonstrate the tunability between the two extremes, that is, field quadrature and photon number measurements. The displacement of the dips of the Wigner functions from the origin confirms the phase-sensitivity of the POVM elements, as in Figs.~(\ref{Fig:homo_APD} d-f). This displacement is given by~\cite{Laiho_ACS09}
\begin{equation}
\mathcal{D} = \sqrt{\frac{(1-T)\mathcal{M}}{T}}|\alpha_{LO}|,
\label{eq:displacement}
\end{equation}
matching our measurement results, which go up to 1.8, about 3.6 times the quadrature uncertainties of coherent states. Futher displacement can be achieved with higher LO amplitude. Allied with the phase adjustment, this tunable region is sufficient to probe most of the non-classical state generated to date~\cite{Ourjoumtsev07, Laiho_CGS10, Gerrits10}. The Wigner functions in Figs.~(\ref{Fig:homo_APD} d, e) have negative values, which clearly demonstrate the non-classicality of these detectors. In contrast, the Wigner function in Fig. \ref{Fig:homo_APD} (f) is positive everywhere. The loss of the quantum feature is due to the increased mode-mismatch, which manifests itself as decoherence due to the two competing incoherent processes stated above. These results reiterate the importance of QDT when quantum detectors become complex; it allows us to account for the external degrees of freedom which are difficult to control and may change the detector response in an unexpected way.

Finally, we apply QDT to a weak-homodyne PNRD. Such a detector has been predicted to be very powerful for non-classical state preparation and measurement~\cite{Puentes09}. It has been experimentally used in the probe of a non-classical Wigner function in a point by point manner~\cite{Laiho_CGS10}, and a demonstration of sensitivity beyond the standard quantum limit of coherent optical communication~\cite{Kenji11}. For the PNRD in our setup, a time-multiplexed detector with $N=9$ time bins~\cite{Achilles_SSBW03,Achilles_SW06}, convolution effects, limited detection efficiency ($24\%$) and interference with the LO require us to extend the Hilbert space to $d=450$ (See Methods: Estimation of $d$). The total number of real parameters involved $(N-1)d^2 \sim\ 1.8 \times 10^6,$  considerably exceeds the largest quantum tomography ever performed, that of an 8-qubit state with 65536 parameters~\cite{8qubit_05}. Our recursive reconstruction method provides a tractable solution to this problem which is infeasible with the standard approach. Figs.~(\ref{Fig:3click} a, b) shows the experimentally reconstructed POVM elements of the 1-click and 3-click events (both displayed up to $d=80$ photons). Their distinctive ranges of sensitivity are evident. The Wigner function of the 1-click POVM element, shown in Fig.~(\ref{Fig:3click} c), has an overlap of $98\%$ with a $\mathcal{D}=1.62$ displaced single-photon state (having experienced $84.3\%$ loss due to the detection efficiency and interference beam splitter). The negativity of the Wigner function is direct evidence of its non-classicality, and professes its suitability for engineering non-classical states.

Phase-sensitive measurements are crucial to fully exploit the fundamental features of quantum physics and to optimally utilize optical telecommunications channels. Weak-field homodyne detectors with photon-number resolution are a unique phase-sensitive measurement in that they respond concurrently to both wave-like and particle-like characteristics of input quantum
states. They hold great potential for applications in quantum information science operating in a hybrid continuous-discrete setting, and fundamental investigations of quantum mechanics.  We used QDT to elucidate the simultaneous wave and particle sensitivity of weak-homodyne photon-number-resolving detection. Our QDT scheme does not rely on the technical details of the measurement process, providing a universal or device-independent understanding of the role of quantum coherence in a measurement process. It foreshadows a new means of assessment and verification of more complex optical detectors, for example, that can attain superadditive capacity and the Holevo limit in coherent communication~\cite{Guha11}. 

\section{Methods}

\textbf{Experimental details:} The local oscillator and probe states are generated by an amplified Ti:Sapphire laser (Coherent Mira Seed, followed by Coherent RegA regenerative amplifier. Operating wavelength $\lambda_0=$ 830 nm at repetition rate $f_R=$ 256.752 kHz). We use a Semrock interference filter with full-width at half-maximum bandwidth of $\Delta\lambda=$ 3 nm to reduce the bandwidth and thus the effects arising from dispersion in the optical elements.

The laser beam is split at a broadband beam splitter (BBS, reflectivity $35\%$) to generate LO and probe beams. The probe and LO then each undergo different control elements, to be finally interfered at another BBS. LO control is performed by a half-wave plate (HWP) and a polarizing beam splitter (PBS), setting the power of the LO, as well as defining its polarization. While the LO strength is fixed for an experimental run, the probe states' amplitude and phase are varied. We use a HWP followed by a Glan-Thompson polarizer (GT) to adjust the amplitude $|\alpha|$ of the probe states with a dynamic range of $10^5$. A beam sampler with low reflectivity sends a fraction of the probe beam to a NIST-traceable Coherent FieldMaxII-TO power meter to monitor the variable attenuation realized by the HWP-GT combination.

We control the phase of the probe state $\theta$ with a variable delay line driven by a piezo translator (Physik Instrumente P-841.30). Moving the delay line changes the phase between LO and probe at the recombination BBS, where LO and probe interfere.

One output of the interference BBS passes through a set of pre-calibrated neutral density (ND) filters, and is coupled into a single-mode fiber to be detected by an APD or time-multiplexed PNRD. The other output of the recombination BBS is sent to a fast photodiode to monitor $\theta$. The mode-overlap $\mathcal{M}$ is also measured with this photodiode by balancing the probe and LO, and calculating the visibility of the interference fringes when $\theta$ is scanned.

For each probe state amplitude $|\alpha|$, we use forty phase settings $\theta$, uniformly distributed between 0 and $2\pi$ and measure the click statistics for $0.5$ second.

\textbf{Derivation of Eq.~(\ref{eq:rela_q_diag}):} Using $\alpha = |\alpha|e^{i\theta},$ the coherent state projector can be expressed as
\ben
\proj{\alpha} &=& e^{-|\alpha|^2}\sum_{r,s}\frac{\alpha^r\alpha^{*s}}{\sqrt{r!s!}}\ket{r}\bra{s}  \nonumber\\
 &=& e^{-|\alpha|^2}\sum_{r,s}\frac{|\alpha|^{r+s}}{\sqrt{r!s!}}e^{i(r-s)\theta}\ket{r}\bra{s}.
\een
Since $Q_n(\alpha) = \tr(\hat{\Pi}_n\proj{\alpha})/\pi,$ with $\hat{\Pi}_{n} = \sum_{j,k}\Pi_{n}^{j,k}\ket{j}\bra{k},$ Eq.~(\ref{eq:rela_q_diag}) follows from $\int_0^{2\pi}e^{i(r-s-l)\theta} d\theta = 2\pi\delta_{r,s+l}.$


\textbf{Reconstruction procedure:} For an $N$-outcome detector, and $p$ coherent states as probes $(\ket{\alpha_1},\cdots,\ket{\alpha_p})$, we have $Np$ linear equations provided by Eq.~(\ref{eq:Born}) $p_{n|m} = \tr(\ket{\alpha_m}\langle\alpha_m |  \hat{\Pi}_{n})$ with $m=1\cdots p$ and $n=1\cdots N$. When the maximum photon number is truncated at $d-1$, the set of linear equations can be rewritten as
\be
P = F\tilde{\Pi},
\ee
where $P_{p\times N}$ consists of the measurement statistics, $F_{p\times d^2}$ has as its rows the coherent state probes with $F_{i,j} = e^{-|\alpha_i|^2}|\alpha_i|^{2j}/j!,$ and $\tilde{\Pi}_{d^2\times N}$ has as its columns the POVM elements that are to be reconstructed (first column has the matrix elements of the 0-click POVM, second column for the 1-click POVM, \textit{etc.}). The physical POVM set consistent with the data can be estimated using the constrained convex optimization
\ben
\min \{||P-F\tilde{\Pi}||_2 &+& g(\tilde{\Pi})\}, \nonumber \\
\mbox{subject to~~~} \hat{\Pi}_n \geq 0, && \sum_{n=0}^{N-1}\hat{\Pi}_n = I,
\label{eq:sdp}
\een
where $||M||_2 = \sqrt{\tr(M^{\dag}M)}$ is the Frobenius or the Hilbert-Schmidt norm, and $g(\tilde{\Pi})$ is the regularization function whose form will be given later.

The reconstruction proceeds recursively, starting with the diagonals of $\hat{\Pi}_n.$ We construct the matrix $P^l$ by averaging the measured statistics as per the left-hand side of Eq.~(\ref{eq:rela_q_diag}). A similar averaging of the input states leads to an input matrix $F^l,$ and satisfies the identity
\be
P^{(l)} = F^{(l)}\tilde{\Pi}^{(l)},
\ee
where $\tilde{\Pi}^{(l)}$ corresponds to only the $l^{\mathrm{th}}$ diagonal of the POVM matrices. For each $l,$ one can then setup an independent semi-definite program as in Eq.~(\ref{eq:sdp}). The constraints however need to be tailored to this new scenario. For $l=0,$ this is trivial, as the diagonal elements must all be positive, and the sum over the POVM elements be the identity 
\be
\Pi_{n}^{j,j} \geq 0 \,\,\,\,\forall n,j; \,\,\,\,\,\,\sum_{n=0}^{N-1} \Pi_{n}^{j,j} =1 \,\,\,\,\forall j.
\ee
This same condition requires that the corresponding off-diagonal elements must all add to zero. The positivity condition is recursively enforced using Sylvester's criterion, which states that a matrix is positive if and only if all of its principal minors are positive. As an example, for $l=1$ the condition is
\be
\left|\Pi_{n}^{j,j+1}\right| \leq \sqrt{\Pi_{n}^{j,j}\Pi_{n}^{j+1,j+1}} \,\,\,\,\forall n,j; \,\,\,\,\,\,\sum_{n=0}^{N-1} \Pi_{n}^{j,j+1} =0 \,\,\,\,\forall j.
\ee
Beginning with a positive diagonal matrix, which we obtain by solving the $l=0$ reconstruction, the successive off-diagonals are added, ensuring at each step that the operator thus obtained is positive.

The reconstruction problem effectively deconvolves a coherent state from the statistics to obtain the POVM set. This is an ill-conditioned problem, as seen by the large ratio between the largest and smallest singular values of the matrix $F.$ This makes the POVM extremely vulnerable to small fluctuations in the statistics. This instability is taken care of by the regularization function $g,$  a convex quadratic function, that still allows us to cast the regularized problem as a semi-definite program~\cite{feito2009measuring}. The same regularization function is enforced for each $l,$  one that penalizes large differences $|\tilde{\Pi}_n^{j,j+l}-\tilde{\Pi}_n^{j+1,j+l+1}|$ as $g(\tilde{\Pi})=\gamma \sum_{j}|\tilde{\Pi}_n^{j,j+l}-\tilde{\Pi}_n^{j+1,j+l+1}|^2.$ Note that the regularization makes no assumption about the details of the quantum detector or the actual value of $\gamma.$ Variations of $\gamma$ over two orders of magnitude produced only around $10\%$ difference in the reconstructed POVM.

\textbf{Estimation of $d$:} The response of the time-multiplexed PNRD can be modeled as a loss matrix followed by a convolution matrix, which accounts for the effect of the beam-splitter network. As in Ref.~\cite{feito2009measuring}, it can be shown that without losses, saturating 8 bins with a probability of $99\%$ requires 52 photons or more (assuming both beam-splitters are 50:50). Limited transmission ($0.157$, the transmission of the interference BS $0.655$ times the efficiency of the time-multiplexed PNRD $0.24$) adds other statistical reduction to the number of photons. To have at least 52 photons survive the losses with a probability of $99\%$ requires more than 438 photons at the input. Finally, taking into account the destructive interference with LO, it requires approximately 450 photons to saturate the weak-homodyne time-multiplexed PNRD.

\section{Acknowledgements}
We thank G. Donati, T. J. Bartley, J. Eisert, X. Yang, A. Feito for assistance and fruitful discussions. This work was funded in part by EPSRC (Grant EP/H03031X/1), US EOARD (Grant 093020), EU Integrated Project Q-ESSENCE and the Alexander von Humboldt Foundation.

\section{Author contributions}

L.Z., H.C-R., A.D. contributed equally to this work. L.Z., H.C-R., X-M.J. and I.A.W. conceived the project, and contributed to the design of the experiment, as well as to laboratory measurements and data analysis. L.Z., A.D. and  M.B.P contributed modeling and data analysis. G.P., J.S.L. and B.J.S. contributed to the initial conception of the project. All authors contributed towards the writing of the manuscript.

\section{Additional information}

The authors declare no competing financial interests. Correspondence and requests for materials should be addressed to LZ at l.zhang1@physics.ox.ac.uk.


\end{document}